\documentclass[twocolumn,trackchanges]{aastex62}

\def\SZ{{\it Suzaku }}

\graphicspath{{./}{figures/}}

\received{December 12, 2020}
\accepted{\today}
\submitjournal{ApJ}

\shorttitle{Origin of Galactic Spurs}
\shortauthors{Kataoka et al.}

\begin{document}

\title{Origin of Galactic Spurs: New Insight from Radio/X-ray All-sky Maps}

\correspondingauthor{Jun Kataoka}
\email{kataoka.jun@waseda.jp}

\author{Jun Kataoka}
\affiliation{Faculty of Science and Engineering, Waseda University, 3-4-1, Okubo, Shinjuku, Tokyo 169-8555, Japan}

\author{Marino Yamamoto}
\affiliation{Faculty of Science and Engineering, Waseda University, 3-4-1, Okubo, Shinjuku, Tokyo 169-8555, Japan}

\author{Yuki Nakamura}
\affiliation{Faculty of Science and Engineering, Waseda University, 3-4-1, Okubo, Shinjuku, Tokyo 169-8555, Japan}

\author{Soichiro Ito}
\affiliation{Faculty of Science and Engineering, Waseda University, 3-4-1, Okubo, Shinjuku, Tokyo 169-8555, Japan}

\author{Yoshiaki Sofue}
\affiliation{Institute of Astronomy, The University of Tokyo, 2-21-2, Osawa, Mitaka-shi, Tokyo 181-0015, Japan}

\author{Yoshiyuki Inoue}
\affiliation{Department of Earth and Space Science, Osaka University, 1-1 Machikaneyamacho, Toyonaka, Osaka, 560-0043, Japan}
\affiliation{Interdisciplinary Theoretical \& Mathematical Science Program (iTHEMS), RIKEN, 2-1 Hirosawa, Saitama 351-0198, Japan}
\affiliation{Kavli Institute for the Physics and Mathematics of the Universe (WPI), UTIAS, The University of Tokyo, Kashiwa, Chiba 277-8583, Japan}

\author{Takeshi Nakamori}
\affiliation{Department of Physics, Faculty of Science, Yamagata University,
  990-8560, Japan}
  
\author{Tomonori Totani}
\affiliation{Department of Astronomy, The University of Tokyo, 7-3-1, Hongo, Bunkyo-ku, Tokyo 113-0033, Japan}



\begin{abstract}
 In this study, we analyze giant Galactic spurs seen in both radio and X-ray all-sky maps to reveal their origins. We discuss two types of giant spurs: one is the brightest diffuse emission near the map's center, which is likely to be related to Fermi bubbles (NPSs/SPSs, north/south polar spurs, respectively), and the other is weaker spurs that coincide positionally with local spiral arms in our Galaxy (LAS, local arm spur). Our analysis finds that the X-ray emissions, not only from the NPS but from the SPS are closer to the Galactic center by $\sim$ 5$^{\circ}$ compared with the corresponding radio emission. Furthermore, larger offsets of 10$-$20 $^{\circ}$ are observed in the LASs; however, they are attributed to different physical origins. Moreover, the temperature of the X-ray emission is $kT$ $\simeq$ 0.2~keV for the LAS, which is systematically lower than those of the NPS and SPS ($kT$ $\simeq$ 0.3~keV) but consistent with the typical temperature of Galactic halo gas. We argue that the radio/X-ray offset and the slightly higher temperature of the NPS/SPS X-ray gas are due to the shock compression/heating of halo gas during a significant Galactic explosion in the past, whereas the enhanced X-ray emission from the LAS may be due to the weak condensation of halo gas in the arm potential or star formation activity without shock heating.
\end{abstract}



\section{Introduction} \label{sec:intro}
 The all-sky survey is a unique albeit the only approach for revealing giant structures in the sky that are rarely seen in pointing observations with a limited field of view (FOV). Haslam et al. (1982) conducted the first complete radio survey, which was measured at 408 Hz; the survey confirmed various giant spurs and loop structures extending over the entire sky. Of particular note was Loop I, a continuum loop spanning across 100$^{\circ}$ in the sky, and its brightest arm, known as the north polar spur (NPS). Initially, it was argued that NPS/Loop I was an
old supernova remnant that was extremely close to the Sun (Berkhuijsen 1971); however, an alternative idea
was proposed, suggesting that it was the remnants of starburst or nuclear outbursts in the Galactic center (GC) that occurred over 10 million years ago (Sofue 1977). Although the latter idea successfully explained similar structures observed in the south (SPS, south polar spur: Sofue et al. 2000), it was almost neglected; however, it received renewed attention after the discovery of Fermi bubbles (Su et al. 2010).

 Fermi bubbles are giant structures extending approximately 50$^{\circ}$ (or 8.5~kpc) above and below the GC, with a longitudinal width of 40$^{\circ}$. Notably, the NPS/Loop I exhibited close contact with Fermi bubbles. Moreover, Fermi bubbles are spatially correlate with the WMAP haze (Dobler \& Finkbeiner 2008) measured between 20$-$50 GHz, which was confirmed later via Planck observations (Planck collaboration 2013). The connection between the NPS/Loop I and Fermi bubbles was widely discussed based on $ROSAT$ all-sky X-ray maps (Snowden et al. 1995), although a positional offset between radio and X-ray spurs was suggested (Sofue et al 2015; Kataoka et al. 2018). Using multiple observations with $Suzaku$, the X-ray emission from the NPS/Loop I was well represented by $kT$ $\simeq$0.3 keV thin thermal plasma (Kataoka et al. 2013); this may be interpreted as a result of shock heating of Galactic halo gas during the explosion, characterized initially by a temperature of $kT$ $\simeq$ 0.2 keV (Yoshino et al. 2008; Henley et al.2010; Nakashima et al. 2018).
More recently, $eROSITA$ on the Spectrum--Roentgen--Gamma mission (Predehl et al. 2020a) launched in June 2019 provided a new, sharp all-sky map, in which a clear X-ray envelope surrounding Fermi bubbles was observed in both the northern and southern skies (Predehl et al. 2020b). Hence, the NPS/Loop I and SPS can now be regarded as the remnant of a Galactic explosion over 10 million years ago.

\begin{deluxetable*}{lcc}[t!]
\tabletypesize{\scriptsize}
\tablecaption{Regions for correlation analysis.}
\tablewidth{0pt}
\tablehead{
  \colhead{Source Region} & \colhead{Galactic longitude} & \colhead{Galactic latitude} \\
}
\startdata
NPS & 10$^{\circ}$ $\le$ $l$ $\le$ 40$^{\circ}$  & 15$^{\circ}$ $\le$ $b$ $\le$ 80$^{\circ}$  \\
SPS & 295$^{\circ}$ $\le$ $l$ $\le$ 335$^{\circ}$  & $-$60$^{\circ}$ $\le$ $b$ $\le$ $-$15$^{\circ}$ \\
LAS1 & 60$^{\circ}$ $\le$ $l$ $\le$ 95$^{\circ}$  & 15$^{\circ}$ $\le$ $b$ $\le$ 35$^{\circ}$  \\
LAS2 & 240$^{\circ}$ $\le$ $l$ $\le$ 280$^{\circ}$  & 15$^{\circ}$ $\le$ $b$ $\le$ 35$^{\circ}$  \\
\enddata 
\end{deluxetable*}

 In addition to the abovementioned giant structures, which are likely located in the GC, weaker but evident spurs were observed in all-sky radio maps. In fact, nonthermal spiral-arm emission along the Galactic plane
is closely associated with the discrete spiral pattern of the Galaxy (Mills 1959; Sofue 1976; see also Nakanishi \& Sofue (2006) for H~I and H$_2$ patterns). Moreover, diffuse radio emissions emanating vertically from the Galactic plane are noteworthy; they are the most conspicuous at longitudes of $l$ $\sim$80$^{\circ}$ and $\sim$260$^{\circ}$, i.e., exactly at the position of the Orion--Cygnus arms (or local arm), where our Sun is located. It has been argued that radio emissions associated with local arm spurs (LASs) are enhanced in nonthermal banks, which extend vertically up to $z$ $\sim$1 kpc from the Galactic plane (Sofue 1973; 1976; 2000). In this model, nonthermal banks are generated by inflations of magnetic fields with cosmic rays as a result of the Parker instability (Parker 1966; 1969); hence, the inflation enhanced above gaseous spiral arms associated with Galactic shocked regions (Fujimoto 1966; Roberts 1969). In X-rays, similar diffuse emissions associated with LASs are visible in both $ROSAT$ and $eROSITA$ all-sky maps; however, their characteristics are yet to be elucidated. In this context, diffuse X-ray emission along spiral arms, whose morphology matched well with those in the mid-infrared or H${\alpha}$ region, was discovered in nearby spiral galaxies by recent $Chandra$ observations (Tyler et al. 2004; Long et al. 2014).

Herein, we provide systematic comparisons of giant Galactic spurs observed in radio and X-ray all-sky maps to reveal their origins. The remainder of this paper is organized as follows. In $\S$2, we present the method and result of correlation analysis to quantitatively discuss the positional offset between the radio and X-ray maps for all Galactic spurs. Next, we analyze the archival \SZ data to reveal the origin of thermal X-ray emission associated with SPS and LAS regions; subsequently, we compare the results with those reported in the NPS/Loop I (Kataoka et al. 2013; 2018). In $\S$3, we discuss the different characteristics of thermal emission in the NPS/SPS and LAS. We first discuss the origin of the comparable thickness and offset between the NPS and SPS; subsequently, we consider the origin of radio and X-ray emissions associated with LASs. Finally, a brief summary and future prospects are presented in $\S$4.

\section{Analysis and Results} \label{sec:ana}
\subsection{Correlation analysis of radio and X-ray all-sky maps}
 We first compare the radio sky map obtained at 408~MHz (Haslam et al. 1982: contour) and the X-ray map measured at 0.75~keV based on the $ROSAT$ all-sky survey (Snowden et al. 1995: color), as shown in Figure 1.  The Galactic spurs analyzed herein are tagged $yellow$, with the projection example for each rectangular region shown separately in the top panels. A significant offset was observed between the radio and X-ray profiles, at least for a certain fixed latitude $b$, as shown in Figure 1. Subsequently, we created a complete list of the radio/X-ray pixel values at the same Galactic position, ($l$, $b$), with a 1$^{\circ}$ resolution in both the longitude and latitude directions. Next, we defined the source extraction region corresponding to the NPS, SPS, LAS1, and LAS2,
as summarized in Table1. Figure~2 shows the close-up view of the radio/X-ray maps for each region.

\begin{deluxetable*}{lllllllllll}[t!]
\tabletypesize{\scriptsize}
\tablecaption{\SZ observations and analysis results}
\tablewidth{0pt}
\tablehead{
  \colhead{SrcID} & \colhead{ObsID$^a$} & \colhead{RA$^b$} &  \colhead{DEC$^b$}  & \colhead{$l$$^b$} & \colhead{$b$$^b$} &  \colhead{Exposure$^c$} & $N_{\rm H}/N_{\rm H,Gal}$$^d$  & \colhead{$kT^e$} & \colhead{EM$^f$} & \colhead{$\chi^2$/~d.o.f$^g$}\\
\colhead{} & \colhead{} & \colhead{[$^{\circ}$]} & \colhead{[$^{\circ}$]} & \colhead{[$^{\circ}$]} & \colhead{[$^{\circ}$]} &  \colhead{[ksec]} & & \colhead{[keV]} & \colhead{[$\times$$10^{-2}$~cm$^{-6}$pc]} & \colhead{}
}
\startdata
\multicolumn{11}{c}{South Polar Spur} \\
\tableline
S1 &   705013010  & 265.961 & $-$76.342 & 317.082 & $-$22.458  & 42.4 & $<$ 0.61 & 0.277$^{+0.029}_{-0.023}$ & 0.62$^{+0.14}_{-0.12}$ & 234.0/231 \\
S2 &   701052010 & 292.839  & $-$72.655  & 322.501 & $-$28.769  & 113.5 & 7.0$-$10.8  & 0.301$^{+0.004}_{-0.003}$ & 4.19$\pm$0.10 & 602.2/418 \\  
S3 &   806082010  & 321.058 & $-$63.430 & 330.620 & $-$40.844  & 36.3 & 2.3$-$2.8 & 0.272$^{+0.013}_{-0.020}$ & 2.09$^{+0.42}_{-0.22}$ & 210.9/160 \\
S4 &   806079010  & 319.721  & $-$63.575 &   330.734 & $-$40.236  & 69.5 & 1.0$-$5.0  & 0.292$^{+0.011}_{-0.010}$ & 1.81$\pm$0.14 & 444.1/312  \\  
\tableline
\multicolumn{11}{c}{Local Arm Spur 1} \\
\tableline
L1 &   704008010 &265.005 & 52.168  & 79.523 & 31.837  & 22.5 & 2.4$-$6.7  & 0.196$^{+0.025}_{-0.014}$ & 2.77$^{+0.97}_{-1.03}$  & 111.2/90\\
L2 &   403008010  & 282.080 & 47.990 &  77.411 & 20.299  & 44.4 & $<$1.4 & 0.204$^{+0.029}_{-0.020}$ & 1.20$^{+0.61}_{-0.40}$  & 299.4/256\\
L3 &   704051010 & 260.921  & 55.895 & 83.893 & 34.333  & 35.9 & 5.8$-$14.8 & 0.217$\pm$0.042 & 0.60$^{+0.77}_{-0.24}$ & 139.9/166\\
L4 &   100017010  & 279.235 & 38.784 & 67.448 & 19.237  & 11.4 & $<$ 2.6  & 0.229$^{+0.014}_{-0.011}$ & 1.83$\pm$0.27 & 147.2/104 \\
L5 &   707038010  & 274.923 & 45.536 & 73.363 & 24.300  & 30.9 & $<$ 1.3 &  0.266$^{+0.018}_{-0.023}$ & 1.52$^{+0.39}_{-0.22}$  & 183.3/168\\
\tableline
\multicolumn{11}{c}{Local Arm Spur 2} \\
\tableline
L6 &   807071010  & 160.083 & $-$35.328  & 274.839 & 20.293  & 45.7 & $<$ 1.0  & 0.229$^{+0.008}_{-0.007}$ & 3.66$\pm$0.35 & 304.7/258 \\
L7 &  808063010 & 163.275 &  $-$40.329  & 279.735 & 17.153  & 152.2 & 3.3$-$4.4 &  0.197$^{+0.011}_{-0.008}$ & 2.97$\pm$0.51 & 801.3/585\\
\enddata 
\tablecomments{$^a$: \SZ observation ID.\\
$^b$: Right Ascension, Declination, Galactic Longitude, Galactic Latitude  of
  \SZ observations.\\
$^c$: \SZ XIS exposure in ksec.\\
$^d$: The ratio of absorbing column density to the full Galactic column along the line of sight when $N_{\rm H}$ was left free in the spectral fitting.\\
$^e$: Temperature of the Galactic halo gas fitted with the \textsc{apec} model for the fixed abundance $Z$ = 0.2$Z_{\odot}$.\\  
$^f$: Emission measure of the Galactic halo gas fitted with the \textsc{apec} model for the fixed abundance $Z$ = 0.2$Z_{\odot}$.\\
  $^g$: $\chi^2$ of the spectral fitting to the model \textsc{apec1+wabs*(apec2+pl)}, where we fixed $kT$ at 0.1~keV, and assumed a Solar abundance, $Z_{\odot}$, for \textsc{apec1}. A photon indexof \textsc{pl} representing the Cosmic X-ray background (CXB) is 1.41 (Kushino et al. 2002). Since \textsc{wabs} is not well constrained due to low photon statics, we fixed them to zero for the values listed here.\\}
\end{deluxetable*}

\begin{figure*}[t!]
\begin{center}
  \includegraphics[keepaspectratio,width=\linewidth]{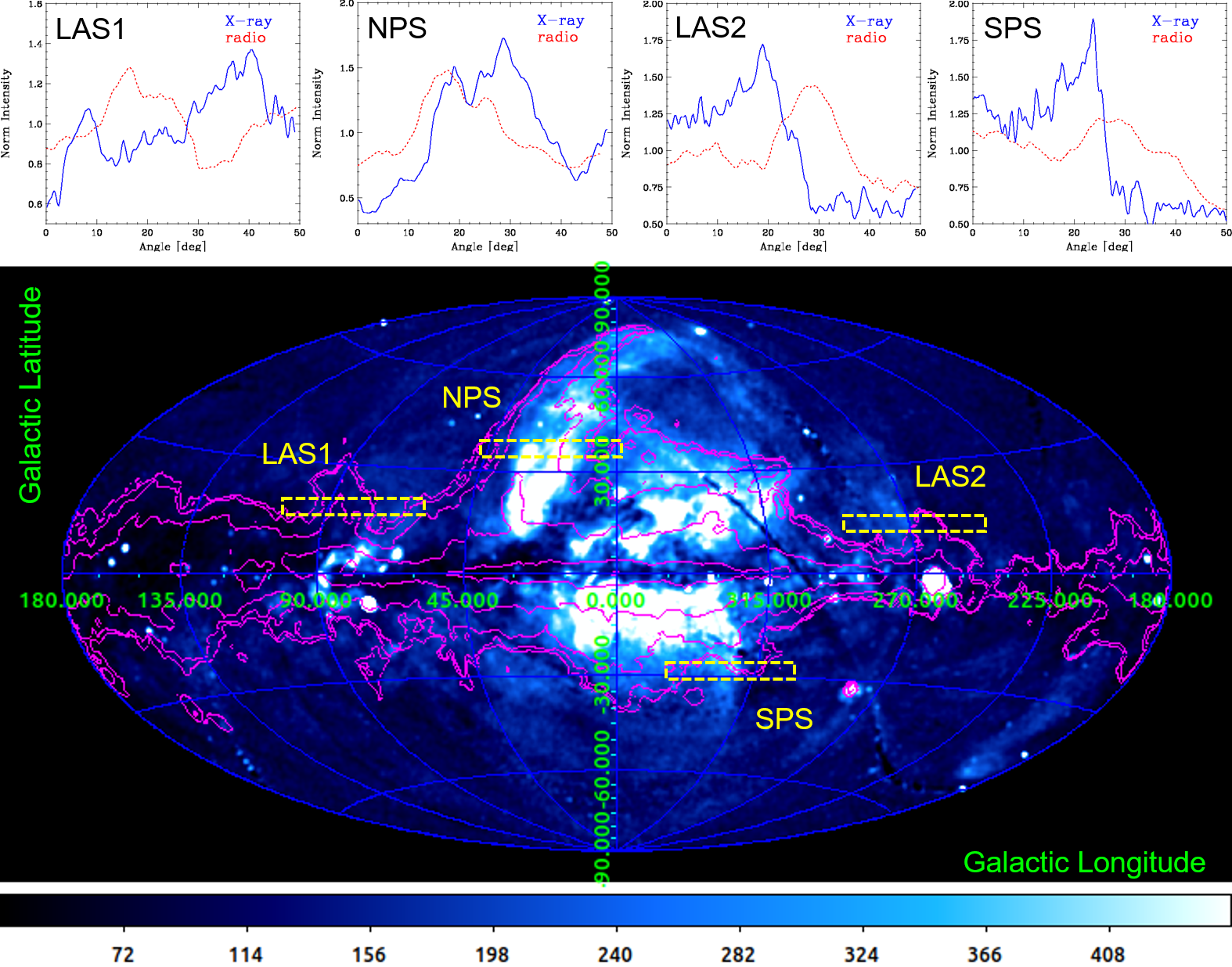}
  \caption{
Galactic spurs analyzed in this paper. NPS: North polar spur,
    SPS: South polar spur, LAS1: local arm spur at $l$ $\sim$ 80 $^{\circ}$,
    LAS2: local arm spur at $l$ $\sim$ 260$^{\circ}$.
    X-ray: 0.75~keV $ROSAT$ all-sky map is shown in logarithmic color
    in the unit of 10$^{-6}$ cts s$^{-1}$ arcmin$^{-2}$.
    Radio : 408~MHz all-sky map is shown as 6-level contours ($magenta$)
    from 30 [K] to 600 [K] in logarithmic scale.
    Projections of radio/X-ray images along the Galactic longitude
    for each rectangular regions ($yellow$: approximately
    50$^{\circ}$$\times$5$^{\circ}$ in longitude and latitude directions)
    are shown in the upper panels,
    where intensities are normalized with each mean values.     
  }
\label{fig:allsky}
\end{center}
\end{figure*}

 To be more precise, we first calculated the correlation coefficients $\rho$ between all radio and X-ray pixel values, i.e., $R(l,b)$ and $X(l,b)$, for each spur region. We defined $\rho$[0] for raw radio and X-ray data as follows:
\begin{equation}
  \rho[0] = \frac{cov(R,X)}{\sigma_{R}~\sigma_X},  
\end{equation}  
where $cov(R,X)$ is the covariance; $\sigma_R$ and $\sigma_X$ are the standard deviations of variables $R$ and $X$, respectively. Next, we calculated $\rho$[$\theta$], which is the correlation coefficient between $R$($l$+$\theta$,$b$) and $X$($l$,$b$), wherein the radio data were shifted by $\theta$ [$^{\circ}$] in the longitude direction. In this context, a ``positive'' offset is defined as the direction to which the data are shifted toward the center of the Galactic east region (0$^{\circ}$ $<$ $l$ $<$ 180$^{\circ}$), and toward the anti-GC for the west region (180$^{\circ}$ $<$ $l$ $<$ 360$^{\circ}$).

\begin{figure*}[t!]
\begin{center}
  \includegraphics[keepaspectratio,width=\linewidth]{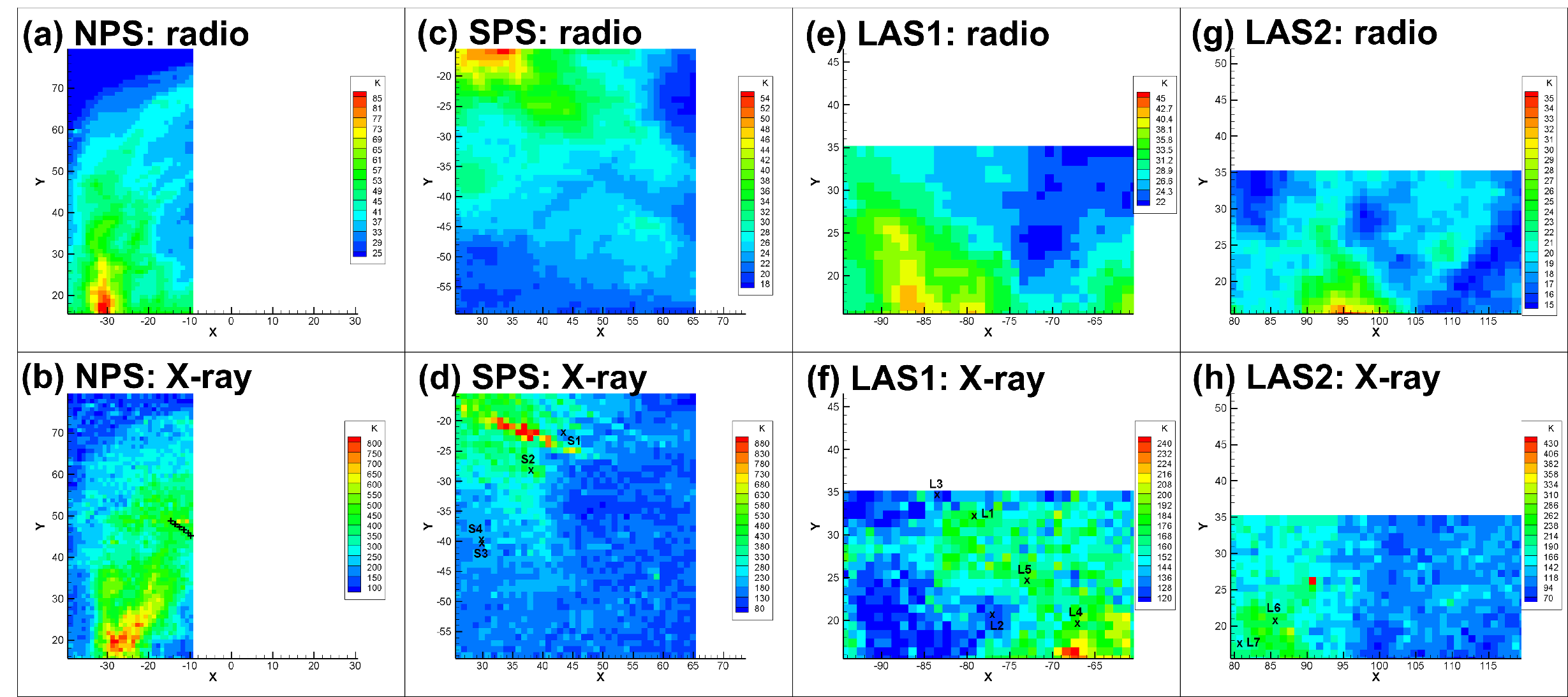}
  \caption{
    Close-up of Galactic spurs as observed from $ROSAT (0.75~keV)$ and HASLAM (408~MHz) all-sky maps. (a) NPS: radio, (b) NPS: X-ray, (c) SPS: radio, (d) SPS: X-ray, (e) LAS1: radio, (f) LAS1: X-ray, (g) LAS2: radio, and (h) LAS2: X-ray. X-ray map ($ROSAT$ 0.75~keV) is shown in units of 10$^{-6}$ cts s$^{-1}$ arcmin$^{-2}$. Radio map (Haslam 408~MHz) is shown in units of Kelvin. Cross (X) indicates pointing center of \SZ observations analyzed herein (see Table.1). In (b), \SZ pointing center for NPS (N1--6 of Kataoka et al. 2013) is shown as ``+.''
  }
\label{fig:closeup}
\end{center}
\end{figure*}

 Figure~3 ($left$) shows $\rho$[$\theta$] as a function of $\theta$ for the NPS, SPS, LAS1, and LAS2, separately. Is it noteworthy that correlations were stronger for the NPS and SPS than for LAS1 and LAS2. Although we did not exclude point sources from all-sky X-ray maps, no such strong sources existed that affected the results of the correlation analysis presented herein. The amount of peak offset is summarized in Figure 3 ($right$) against Galactic longitude $l$. The errors in $\theta$ were estimated by approximating $\rho$[$\theta$] using a single Gaussian function around the peak; hence, the uncertainties of peak positions were estimated in Figure 3 ($right$) . The positional offsets between the radio and X-rays were $\theta$ = 6.31$\pm$0.34$^{\circ}$ for the NPS and $\theta$ = $-$4.17$\pm$0.83$^{\circ}$ for the SPS. Meanwhile, the offsets of the LASs were significantly larger, i.e., $\theta$ = 19.35$\pm$0.17$^{\circ}$ for LAS1 and $\theta$ = $-$10.57$\pm$0.21$^{\circ}$ for LAS2, respectively. We conclude that in all spurs, the X-ray emission is shifted toward the GC direction compared with the corresponding radio emission by 5$-$20$^{\circ}$.

 It is noteworthy that the observed X-ray all-sky map is generally modified by absorption owing to the interstellar medium (mostly H~I gas), particularly in the low latitude regions within or near the Galactic disk and bulge. The amount of absorption depended on the distance to the structure of interest, which is however unknown for the NPS, SPS, and LASs analyzed in this study. Hence, we did not consider such modifications for the correlation analysis in this section.
For this reason, we constrained the analysis regions to a relatively high Galactic latitude of $|$$b$$|$ $>$ 15$^{\circ}$ (see Table 1), where the absorption was almost negligible.

\subsection{Thermal emission from SPS and LAS}
 The X-ray emission properties of the NPS/Loop I have been summarized and discussed in our previous papers (Kataoka et al. 2018 and reference therein). Similarly, in this study, we first investigated all the archival $Suzaku$ data whose pointings were situated in the SPS, LAS1, and LAS2 regions, as listed in Table~2. We excluded pointings that contained either bright X-ray
sources or extended sources such as Galaxy clusters, whose tailed emission might affect the analysis in the same FOV. Consequently, we discovered that four, five, and two pointings were used in the SPS, LAS1, and LAS2 regions, respectively. The pointing centers of the \SZ observations and exposure in kiloseconds are listed in Table~2. The analysis procedure used was the same as those provided in the literature; hence, the results of \SZ observations in the NPS regions (six points in the region defined in Table~1) were reproduced from the data by Kataoka et al. (2013).

 In summary, we extracted the XIS data from XIS~0, 1, and 3, where XIS~0 and 3 were front-illuminated CCDs (FI-CCDs), and XIS~1 was a back-illuminated CCD (BI-CCD) that possessed better sensitivity than FI-CCDs below 1 keV but less sensitive above 5 keV. We analyzed the $Suzaku$ data using $\textsc{headas}$ software version 6.22, $\textsc{xspec}$ version 12.9
and a calibration database released in April 2016. We applied $\textsc{sisclean}$ to remove hot and/or flickering pixels. Only data with a cutoff rigidity larger than 6 GV as well as
day and night Earth data with an elevation angle larger than 20$^{\circ}$ were used for the analysis. We extracted the spectrum after removing possible point sources within the same FOV and then generated redistribution matrix files and auxiliary response files using $\textsc{xisrmfgen}$ and $\textsc{xissimarfgen}$ (Ishisaki et al. 2007). The non-X-ray background spectra from the night Earth observations were generated with $\textsc{xisnxbgen}$.

  For the spectral analysis, we used the 0.5$-$7.0~keV data for XIS~0 and 3, and 0.4$-$5.0~keV data for XIS~1. We fitted all the spectra using $\textsc{xspec}$ with a model comprising three plasma components, similar to previous studies: $\textsc{apec1 + wabs * (apec2 + pl)}$. The $\textsc{apec}$ model assumes an emission from collisionally ionized diffuse gas. Here, (1) $\textsc{apec1}$ is an unabsorbed thermal plasma with $kT$ = 0.1~keV that mimics the local hot bubble (LHB), (2) $\textsc{wabs*apec2}$ is the absorbed thermal plasma emitted from the SPS or LAS, and (3)$\textsc{wabs*pl}$ is an absorbed power-law component
representing the cosmic X-ray background (CXB). We assumed metal abundance $Z$ = $Z_{\odot}$ for the LHB, whereas $Z$ = 0.2 $Z_{\odot}$ was assumed for the SPS and LAS, respectively (see Kataoka et al. 2013). For the CXB, we fixed the photon index $\Gamma$
= 1.41, as determined by Kushino et al. (2002).

 Table~2 lists the spectral fitting parameters for each analysis region, focusing particularly on the temperature $kT$ and emission measure (EM) of plasma component (2). We first fitted the data with the absorption column density ($N_{\rm H}$) as a free parameter, but the results were not constrained well owing to low photon statistics. Hence, we investigated the following two extreme cases: (1) $N_{\rm H}$ was fixed at the Galactic value provided by Dickey \& Lockman et al. (1990); (2) $N_{\rm H}$ was fixed at zero. In each case, the resulting $kT$ value coincided within uncertainties. The reduced $\chi^2$ values for the fitting model were listed with the number of degrees of freedom. Figure~4 ($left$) shows the variations in $kT$ in $\textsc{apec2}$ along the Galactic longitude, whereas Figure 4 ($right$) shows the scatter plots of $kT$ and EM for each Galactic spur. The analysis results in the NPS region ($open$ $magenta$) were from N1 to N6 of Kataoka et al. (2013). It was clear that the $kT$ of the SPS regions was generally $\simeq$0.3~keV, which was higher than those in LAS1 and LAS2 concentrated around $\simeq$0.2~keV, thereby suggesting different emission process origins.

 \section{Discussion} \label{sec:dis}
\subsection{Difference of $kT$ between NPS/SPS and LAS}
 In the previous section, we showed that the diffuse X-ray emission associated with the SPS and LAS was generally reproduced well by a thin plasma model assuming collisional ionized gas, but the temperature in the SPS was slightly higher than that in the LAS. Interestingly, the observed $kT$ $\simeq$ 0.3~keV in the SPS was consistent with those reported in the NPS/Loop I and Fermi bubble edges (Kataoka et al 2018). These results suggested the same physical origin between the NPS and SPS, i.e., the shock heating of the Galactic halo gas through the significant Galactic explosion in the past. This idea was further supported by the recent $eROSITA$ observations, which clearly exhibited X-ray bubbles with sharp edges, both in the
north and south of the GC surrounding the Fermi bubbles, where the SPS was situated at the lowest edge of the south bubble (Predehl et al. 2020b).

  By contrast, $kT$ $\simeq$ 0.2~keV observed in LAS1 and LAS2 were noteworthy because the temperature was typical of that generally observed in the Galactic halo gas (Yoshino et al. 2008; Nakashima et al. 2018), suggesting that the emissions were from local, unshocked halo gas. It was argued that each spiral arm rendered the local potential minimum on the order of 10$\%$ of the gravitational field of the Galaxy; hence, the interstellar gas will fall into the minimum with a typical velocity of 20$-$30 [km s$^{-1}$] (Roberts 1969;
Sofue 1973). It is noteworthy that the sound velocity of the interstellar gas is expressed as follows:
\begin{equation}
c_s = \sqrt{\gamma k_B T/\mu m_p} \simeq  15\left( \frac{T}{10^4 K} \right)^{1/2} {\rm km~ s^{-1}}, 
\end{equation}  
where $\mu$ $\simeq$ 0.61 is the mean molecular weight, and $T$ $\sim$ 10$^4$ [K] is the temperature of the interstellar gas. Hence, the falling gas causes a strong shock of $M$ $\sim$ 2$-$3, where $M$ is the Mach number, resulting in a significant increase in the star formation rate because the compression of the gas density scales as $n$ $\propto$ $M^2$. Moreover,
if the Galactic shock is isothermal and the magnetic field $B$ is parallel to the arm, $B$ $\propto$ $M^2$, then the radio emissivity scales as $\propto$ $n$$B^2$ $\propto$ $M^6$. Therefore, the local arms will be sufficiently bright in the radio map, as first confirmed by Mills (1959).

 Meanwhile, the X-ray emitting gas had a sound velocity of $c_s$ $\sim$200 km s$^{-1}$ for $kT$ $\simeq$ 0.2 keV; hence, it remained unshocked. Even in such a high-velocity gas, an increasing density reflecting the local gravitational potential may occur, although detailed simulations/observations are necessary. Such a gentle pile-up of X-ray emitting gas may 
account for the observed enhancement around the LAS. Because the EM of thin thermal X-ray gas scales as EM $\propto$ $n^2$, only a 20$-$30$\%$ increase in X-ray gas density, $n$, is required to enhance the X-ray gas, whose temperature
is $kT$ $\simeq$ 0.2~keV, as indicated from the observation.

\subsection{Thickness and offset of NPS/SPS}
We observed a significant offset between the radio and X-ray intensity distributions not only in the NPS but in the SPS and LAS structures, although their origins are likely to be different, as indicated in the previous section. For the NPS/SPS, the situation can be interpreted based on the diffusive shock process. The relativistic electrons accelerated via the forward-shock emit synchrotron radiation in the radio band, whereas swept-up, compressed gas emitted thin-thermal X-ray radiation in the reverse-shocked region. In this context, the observed offset between radio and X-rays might be related to the thickness and positional offset of the forward/reverse shock regions. In fact, X-ray emitting shells are typically observed on the inner side of the radio shell in the case of supernova remnants (e.g., Gaetz et al. 2000 for the case of E0102-72.3).

\begin{figure*}[t!]
\begin{center}
  \includegraphics[keepaspectratio,width=0.48\linewidth]{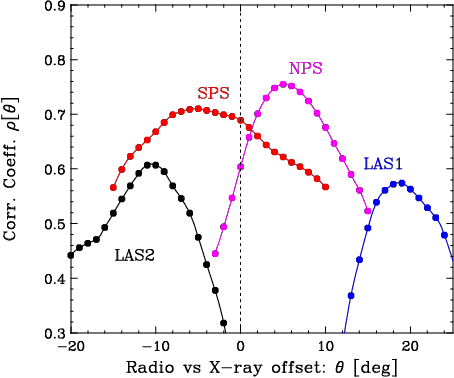}
  \includegraphics[keepaspectratio,width=0.48\linewidth]{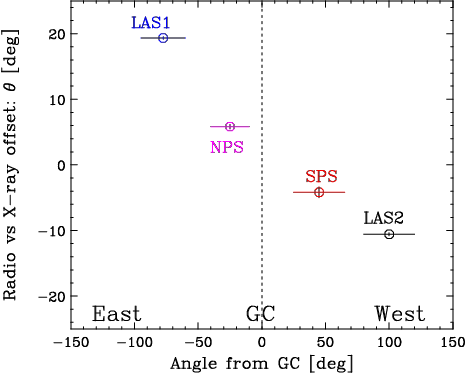}  
  \caption{($left$)
  Correlation coefficient calculated between radio (408~MHz) and X-ray (0.75~keV) spurs, as defined in Figure 2. Angle in horizontal axis becomes positive when X-ray structure is offset toward GC for spurs in Galactic east, whereas negative for spurs in the west. ($right$) Positional offset of each Galactic spur plotted against Galactic longitude. In all cases, X-ray spurs are located on inner side of corresponding radio spurs.  
}
\label{fig:offset}
\end{center}
\end{figure*}

From Kataoka et al. (2015) we adopt a simple model in which two spherical bubbles that mimic the north and south are embedded in the center of a gaseous halo. The radius of the outer shell corresponding to the NPS/SPS is $R$ $\simeq$ 5 kpc and the shock velocity is $v_{sh}$ $\simeq$ 300 km s$^{-1}$, as indicated by X-ray observations. First, the dynamical time scale $t_{dyn}$ in which the bubble expands to radius $R$ is expressed as $t_{dyn}$ $\simeq$ $R/v_{sh}$.
For example, $t_{dyn}$ is 16 Myr in case of $R$ = 5 kpc and $v_{sh}$ = 300 km s$^{-1}$.

 The thickness of forward shock (FS) region can be approximated as follows:
\begin{equation}
d_{FS} \simeq (\frac{k_1}{u_1}+\frac{k_2}{u_2})
  \simeq \frac{5}{v_s}\frac{\xi r_g c}{3}, 
\end{equation}
where $k_1$ and $k_2$ are the diffusion coefficients; $u_1$ and $u_2$ are the plasma velocities in the upstream/downstream of the FS, respectively; $v_s$ is the shock velocity; $r_g$ is the gyro radius; $\xi$ ($>$1) is a constant factor (Drury 1983). Furthermore, we assume that the diffusion coefficient $k_1$ = $k_2$ $\simeq$ $\xi$ $r_g$$c$/3. In addition, $v_s$= $u_1$ = 4$u_2$, $r_g$ = $\gamma$$m_ec^2$/$eB$, where $\gamma$ is the Lorentz factor of the accelerated electrons. The acceleration time scale of electrons with energy $\gamma$$m_ec^2$ is expressed as 
\begin{equation}
t_{acc} =  \frac{20\xi r_g c}{3 u_1^2} = {\frac{4d_{FS}}{v_{sh}}}. 
\end{equation}
For comparison, the radiative cooling time of electrons emitting 408 MHz of radio emission is expressed as
\begin{equation}
  t_{cool,R} = \frac{3m_e c}{4 \sigma_T U_B \gamma}
 \simeq 20 \left(\frac{B}{15{\rm~\mu G}}\right)^{-3/2}~\rm{Myr}, 
\end{equation}
where $m_e$ is the rest mass of electron, $\sigma_T$ = 6.65$\times$10$^{-25}$ cm$^{-2}$ is the Thomson cross section, and $U_B$ is the magnetic field density.
We assume that the shock compression ratio of the magnetic field is $\simeq$ 4. Hence, the cooling time is slightly longer but almost comparable to $t_{dyn}$. Therefore, the thickness of the radio shell, $d_R$, is approximated by equating $t_{acc}$ and $t_{dyn}$ $\simeq$ $t_{cool,R}$, resulting in $d_R$ $\simeq$ $d_{FS}$ $\simeq$ 1/4~$R$. This is on the order
of ~ 1~kpc, which is consistent with the observed thickness of the radio emission region of the NPS/SPS.

Next, we consider the dynamics of the X-ray shell. First, the radiative cooling time scale of the X-ray emitting gas can be expressed as
 \begin{equation}
   t_{cool, X} \simeq \frac{1.1\times10^5~T_6^{1.7}}{n~{\rm cm^{-3}}}~{\rm yr}\\
 \hspace{3mm}{\rm for}\hspace{2mm} 10^5~{\rm K } < T < 10^{7.3}~{\rm K}, 
 \end{equation}
where $n$ is the number density of the X-ray emitting gas, $T_6$ is the
temperature in the unit of $10^6$ K (Eq. 34.4 of Draine (2011)).
We further modified the above equation assuming subsolar metallicity $Z$ $\simeq$ 0.2 $Z_{\odot}$ (Kataoka et al. 2013), which leads to 
 \begin{equation}
   t_{cool, X} \simeq  50 \left (\frac{v_{sh}}{300~{\rm km~s^{-1}}}\right)^{3.4}\left(\frac{n}{0.01~\rm{cm^3}}\right)^{-1} ~\rm{Myr}, 
 \end{equation}  
 (see also Inoue et al. (2017)). We assume a  The X-ray-emitting gas does not cool during the expansion time, $t_{dyn}$; hence, the thickness of the X-ray shell is determined by the amount of halo gas piled up in the reverse shock (RS) region to form the NPS.

  As the underlying halo gas density profile, we assume a hydrostatic isothermal model expressed as
\begin{equation}
n(r) = n_0 \left[1 + \left(\frac{r}{r_c}\right)^2\right]^{-1}, 
\end{equation}  
where $n(r)$ is the gas density [cm$^{-3}$] at radius $r$ from the GC; $n_0$ is the density at $r$ = 0; $r_c$ is the core radius, which we set as $r_c$ = 0.5~kpc from the observation (Kataoka et al. 2015). The thickness of the X-ray shell, $d_X$, can be determined from the conservation of the swept gas mass as follows: 
\begin{equation}
\int^{R}_0 4 \pi r^2 n(r) dr =  4 \pi R^2 [4 n(R)] d_X, 
\end{equation}  
where 4$n(R)$ is the density of the shocked halo gas in the downstream, assuming the strong shock of a specific heat ratio of 5/3. Using Eq. (8), we obtained $d_X$ $\sim$ 1/4~$R$ $\sim$ 1.3 kpc for $R$ = 5 kpc. This is consistent with the observed thickness of the NPS/SPS,
as shown in the X-ray map, and explains the approximate similarity between the radio/X-ray thicknesses,  $d_{R}$ $\sim$ $d_{X}$. Hence, the observed offset of $\sim$5$^{\circ}$, which corresponds to ~0.7~ kpc between the radio and X-ray NPS, may indicate that the FS is located outside the RS by 0.7~kpc on average; however, approximately half of the FS/RS regions may have been overlapped.
This is because the radio emission becomes maximum around the forefront of the FS owing to compressed magnetic field via the synchrotron emission, whereas the X-ray emission would have a broad maximum in the RS reflecting a density gradient of the swept up halo gas.

\begin{figure*}[t!]
\begin{center}
  \includegraphics[keepaspectratio,width=0.48\linewidth]{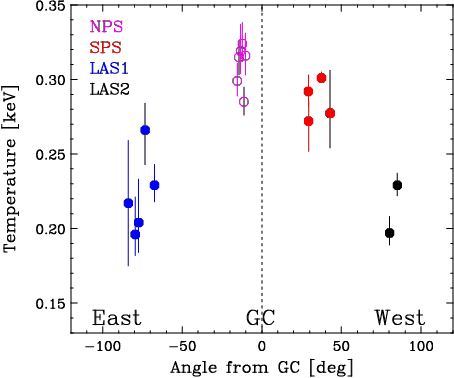}
  \includegraphics[keepaspectratio,width=0.48\linewidth]{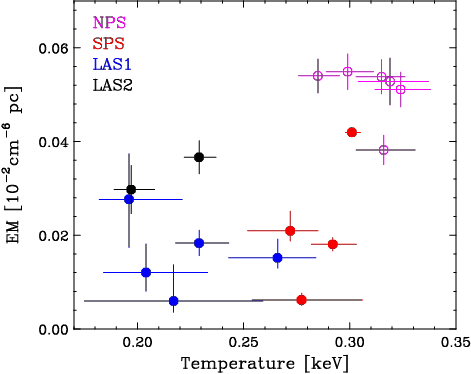}  
  \caption{
    ($left$) Temperature ($kT$ of \textsc{apec2} in Table 2) variation in halo gas along Galactic plane. ($right$) Scatter plot of temperature vs. emission measure of halo gas ($kT$ and EM of \textsc{apec2} in Table 1). Analysis results in NPS region were from N1 to N6 of Kataoka et al. (2013).
}
\label{fig:kTEM}
\end{center}
\end{figure*}

\subsection{Origin of radio/X-ray offset in LAS}
 As reviewed in $\S$1, the radio emissions associated with the LASs have been well known since the 1950s; however, the origin of radio spurs in the tangential direction of the local arm is yet to be elucidated. Sofue (1973;1976) suggested that spurs immediately above the Galactic shock may be generated by the inflation of magnetic fields through the Parker instability triggered by the Galactic shock wave.  Such an inflation is promoted by the strong compression of the gas and magnetic fields; subsequently, the magnetic force lines above the shock lane will be stretched into the halo perpendicular to the Galactic plane. This is consistent with measurements of vertically extended magnetic fields, as inferred from polarized far-infrared dust emissions in the local arm directions (Mathewson \& Ford 1970; Planck Collaboration et al. 2015).

In such a scenario, the formation of radio spurs may be
substantially delayed  after the Galactic shocks are being
activated. First, shock-compressed interstellar gas (mostly H~I and H$_2$),
increases the magnetic field strength
along the arm, and increases the rate of star formation at the shock
front. Thus,  the regions of the
newly born stars (or H~II regions) lies just outside the
shock front. Typical time-scale for the evolution
of massive stars is $t_{SF}$ $\sim$ 10$^{6-7}$ years, thus
the H~II region $lags$ $behind$ the shock by $\simeq$
$t_{SF}$$\times$($v_{\rm rot}$ $-$ $v_{\rm \Omega}$) $\simeq$ 1~kpc along
the direction of
Galactic rotation, where $v_{\rm rot}$ $\simeq$ 220 km s$^{-1}$ is
the rotation speed of the Galaxy and $v_{\rm \Omega}$ $\sim$ 100-150
km s$^{-1}$ is the pattern speed of the spiral density wave.
Accordingly, offset by $\sim$ 1~kpc $\times$
sin($p$) $\sim$ 100~pc is anticipated
in the direction perpendicular to the Galactic arm for a pitch
angle of $p\sim 12^\circ$.
When the stars end their lives,
supernovae happen to accelerate cosmic-ray electrons, which
may take 10$^{4-5}$ years further. Then accelerated electrons
propagate into the nonthermal bank perpendicular to the disk along the arm
within a timescale of $h$/$v_A$ $\simeq$ $\sim$ 10$^6$ yr, where
$h$ $\simeq$ 100~pc is the height of non-thermal bank and
$v_A$ $\simeq$ 10$-$100 km s$^{-1}$ is the Alfven velocity.

 In this context, we can qualitatively interpret the offset between radio and X-rays in the LAS as follows. The position of the radio spur may include an offset compared with the Galactic shock position because we expect a $\sim10^{6-7}$ years of delay for their formation. Radio spurs originate in synchrotron emission by cosmic-ray electrons, which are likely generated by supernova remnants. Hence, time delays due to stellar evolution, particle acceleration, and diffusion should occur. By contrast, the LAS diffuse X-ray emission may be associated with the Galactic shock region if the Galactic halo gas is trapped by the local potential minimum of the spiral arm. Such an association is often observed in nearby galaxies (Tyler et al. 2004; Long et al. 2014). In such cases, the X-ray emission from the LAS is expected to $lead$ the radio emission by $\sim$ a few 10$^6$ years. Assuming a typical distance of $\simeq$ 1~kpc to the regions of the Orion--Cygnus arm, which primarily contributes to the radio/X-ray emissions of the LAS, the observed 10$-$20$^{\circ}$ offset corresponds to $\sim 100$ pc, which is approximately consistent with the horizontal size ($\sim$30$^{\circ}$ from Figure 1)
of the radio bank in the all-sky map, where we assumed $\sim$ 100~pc.

\subsection{Further comments on the NPS}
Finally, we revisit some aspects related to the distance and possible origin of the NPS. 
As noted in \S 1, the NPS was initially thought to be the remnants of an old supernova 
close to the Sun rather than a distant structure associated with the GC activity. 
Although this idea is not well supported by a number of recent X-ray and gamma-ray 
observations, it was recently claimed that the NPS distance between 70 and 135 pc 
depends on the galactic latitudes, according to near-infrared and optical photometry, 
and Gaia DR2 (Das et al. 2020). However, note that Das et al. (2020) observed the extinction 
of stars toward the NPS; thus, they measured the distance to the cold dust (typical temperature 
$T$ $\simeq$ 10$^3$ K) rather than the X-ray brightness of the NPS itself 
($T$ $\simeq$ 10$^{6-7}$ K). In fact, such foreground dust absorption in the Aquila Rift 
is generally used to provide a lower limit on the distance to the NPS (e.g., Sofue 1994, 
2015; Lallement et al. 2016), but it is unlikely that cold dust and the hot X-ray plasma 
of the NPS coexist. Similarly, a wide range of the distance depending on the position in the 
NPS, as proposed by Das et al. (2020), is also unlikely if the NPS is a single 
continuous structure.

However, let us consider the situation if the NPS is really a structure close to the Sun, 
in the context of observed thickness and offset, as discussed in \S 3.2. Now, the radius of 
the NPS is as small as $R$ $\simeq$ 100 pc, and the width of the forward shock, $d_{FS}$, is 
$\simeq$ 1/4 $R$, as anticipated from $t_{dyn}$ $\simeq$ $t_{acc}$ $\simeq$ 0.3~Myr 
(see Eqs. (4) and (5)). Assuming a typical magnetic field strength of $B$ $\simeq$ 10 $\mu$G for an 
old SNR (e.g., Loru et al. 2020), the cooling time of electrons, $t_{cool,R}$, is much longer 
than $t_{acc}$; thus, the accelerated electrons remain uncooled, and the radio-bright NPS 
would be more spherical in shape rather than forming a shell. In contrast, the X-ray shell is 
formed as previously mentioned, i.e., by sweeping up the interstellar medium. We would expect 
quite different morphologies for the radio (sphere-like) and X-ray (shell), which are far from the 
observation. This is an indirect but additional reason supporting why we think the NPS is a 
giant structure associated with the GC.

Finally, note that the most recent observation by HaloSAT enabled coverage 
of the entire bright NPS through 14 observations of approximately 30 ks 
each (LaRocca et al. 2020). These observations provide the first complete 
survey of the NPS thanks to a wide field of view. While the observed X-ray 
spectrum is well fitted by two thermal components of $kT_{cool}$ $\simeq$ 0.1~keV 
and $kT_{hot}$ $\simeq$ 0.3~keV, there is a gradient of temperature across 
the NPS such that the inner arc temperature is slightly warmer 
($kT_{hot}$ $\simeq$ 0.31~keV) than the outer arc ($kT_{hot}$ $\simeq$0.26~keV). 
For comparison, the NPS results presented in Figure 4 ($kT$ = 0.30$\pm$0.01~keV) 
were extracted from Kataoka et al. (2013), in which all \SZ pointings were arranged 
in the inner arc of the NPS. Moreover, LaRocca et al. (2020) provided a new indication 
that the cool component also belongs to the NPS rather than to local hot bubbles (LHB). 
Similarly, even the modeled hot component of $kT$ $\simeq$ 0.30~keV further complicated 
the analysis by adding a halo gas emission of $kT$ $\simeq$ 0.2~keV in addition to 
the NPS (e.g., Gu et al. 2016; Miller \& Bregman 2016). Although the same conclusion 
is always reached, i.e., the NPS is a distant object in the kpc range, 
future deep observations by wide-field satellites such as $HaloSAT$ and $eROSITA$ 
will provide a new insight to resolve such complexity in the modeling of NPS spectra.

 \section{Conclusion} \label{sec:con}

Herein, we analyzed two types of Galactic spurs observed from both radio and X-rays, i.e., the NPS and SPS. These spurs are likely to be the remnants of Galactic explosions in the past and are hence located near the GC, whereas LASs are associated with local spiral arms. The important results of this study are as follows:
\begin{itemize}
{\item the X-ray emission from the NPS and SPS were well represented by a thin thermal plasma of $kT$$\simeq$ 0.3~keV; hence, shock-heated halo gas was most likely the origin.}
{\item the other X-ray emissions from the LAS were represented by a thin thermal plasma of $kT$$\simeq$ 0.2~keV. This can be regarded as the weak condensation of halo gas in the arm potential or star formation activity without shock heating.}
{\item the positional offset observed from the NPS and SPS in the radio and X-ray maps were $\simeq$5$^{\circ}$, where the X-ray emissions were from the inner side of the radio shell. A similar width of radio/X-ray structures as well as positional offsets are naturally interpreted if the radio emission is primarily from the forward shock front, whereas the X-rays were from the downstream of the reverse shock. The thickness of each shock was $\sim$1.3~kpc, half of which was superposed onto each other.}
{\item the positional offset of 10$-$20$^{\circ}$ was observed in the LAS; however, it might be attributed to different origins. Radio spurs might possibly be associated with nonthermal bank, which might have an offset of $\sim$ 100 pc compared with the Galactic shock position. By contrast, the X-ray spurs might be closer to the Galactic shock if the halo gas was simply trapped by the local arm potential or related to the star formation activity in the spiral arm. }
\end{itemize}
We are cognizant that there exist more structures that are possibly related to various Galactic spurs and even to Fermi bubbles' edges, which are marginally visible but not conclusive based on the current HASLAM and $ROSAT$ datasets. 
Moreover, a correlation analysis for the entire Loop-I structure is expected to be conducted, but it is still limited to the brightest eastern part known as the NPS. 
All-sky observations with unprecedented sensitivity and resolution, for example, $Planck$ and $eROSITA$ at different energies, will provide further opportunities to confirm the 
physical origin and structure of Galactic spurs in the near future.

\acknowledgments 

We thank an anonymous referee for his/her constructive comments to improve
this manuscript. This research made use of Astropy, a community-developed core Python package for Astronomy (The Astropy collaboration 2013; 2018). J.K. acknowledges the support from JSPS KAKENHI Grant Numbers JP20K20923.
Y.I. is supported by JSPS KAKENHI Grant Numbers JP18H05458 and JP19K14772.
T.T. was supported by JSPS/MEXT KAKENHI Grant Numbers 18K03692 and 17H06362.



\end{document}